\documentclass[prb, showpacs,twocolumn]{revtex4}
\usepackage{graphicx}
\usepackage{color}

\def\in-situ{{\em in\ situ}}
\def\ex-situ{{\em ex\ situ}}
\def\SRO{\mbox{Sr$_2$RuO$_{4}$}}

\begin{document}

\title{Quantitative analysis of \SRO\ ARPES spectra:\\
Many-body interactions in a model Fermi liquid
}

\author{N.J.C. Ingle}
\altaffiliation[Present address: ]{Department of Physics and Astronomy, University of British Columbia, Vancouver, BC V6T 1Z4, Canada}
\author{ K.M. Shen}
\altaffiliation[Present address: ]{Department of Physics and Astronomy, University of British Columbia, Vancouver, BC V6T 1Z4, Canada}
\author{F. Baumberger}
\affiliation{Department of Physics and Applied Physics, Stanford University, Stanford,
CA 94305}
\author{W. Meevasana}
\affiliation{Department of Physics and Applied Physics, Stanford University, Stanford,
CA 94305}
\author{D.H. Lu}
\affiliation{Department of Physics and Applied Physics, Stanford University, Stanford,
CA 94305}
\author{Z.-X. Shen}
\affiliation{Department of Physics and Applied Physics, Stanford University, Stanford,
CA 94305}

\author{A. Damascelli}
\affiliation{Department of Physics and Astronomy, University of British Columbia, Vancouver, BC V6T 1Z4, Canada}

\author{S. Nakatsuji, Z.Q. Mao, and Y. Maeno}
\affiliation{Department of Physics, Kyoto University, Kyoto 606-8052, Japan and CREST-JST, Kawagushi, Saitama 332-0012, Japan}

\author{T. Kimura, and Y. Tokura}
\affiliation{Department of Applied Physics, University of Tokyo, Tokyo 113-8656, Japan and JRCAT, Tsukuba, 305-0046, Japan}

\date{\today}

\begin{abstract}
ARPES spectra hold a wealth of information about the many-body interactions in a correlated material.  However, the quantitative analysis of ARPES spectra to extract the various coupling parameters in a consistent manner is extremely challenging, even for a model Fermi liquid system.  We propose a fitting procedure which allows quantitative access to the intrinsic lineshape, deconvolved of energy and momentum resolution effects, of the correlated 2-dimensional material \SRO.  For the first time in correlated 2-dimensional materials, we find an ARPES linewidth that is narrower than its binding energy, a key property of quasiparticles within Fermi liquid theory.  We also find that when the electron-electron scattering component is separated from the electron-phonon and impurity scattering terms it decreases with a functional form compatible with Fermi liquid theory as the Fermi energy is approached.  In combination with the previously determined Fermi surface, these results give the first complete picture of a Fermi liquid system via ARPES.  Furthermore, we show that the magnitude of the extracted imaginary part of the self-energy is in remarkable agreement with DC transport measurements.

\end{abstract}

\maketitle

\section{Introduction}

Angle resolved photoemission spectroscopy (ARPES) is an important tool to study electron dynamics, as within the sudden approximation it probes the energy and momentum of the low-energy excitation spectrum of an $N-1$ particle system.  In the non-interacting picture, this spectral function would be a delta function at the precise energy and momentum given by the band structure.  When interactions are turned on, the peak in the spectral function shifts in energy and momentum and gains a finite width that is dependent on the energy and momentum of the excitation.  The lineshape gives direct access to the lifetime of the excitation and can provide insight into the nature of the underlying interactions.  In general, many standard methods used to calculate lifetimes of excitations in correlated systems have starting points which do not allow the inclusion of a full momentum dependent self-energy; they are inherently spatially local descriptions.  Therefore, experimental access to lineshapes via a momentum resolved spectroscopy is of great interest.  This is particularly the case for exotic materials such as the cuprates and manganites, where the strong anisotropy of the elementary excitations might require a description in terms of an explicitly $\bf k$ dependent self-energy.

In general, the shape of measured peaks in an ARPES spectrum does not correspond directly to the excitation's intrinsic lineshape.  This is due to the finite response function, or resolution of the ARPES spectrometer, and possibly also to additional complications from sample surfaces and final state lifetime contributions.  To determine the intrinsic lineshape, all possible effects must be carefully considered when interpreting the experimental spectra.  Therefore, the study of model systems, where these effects can be controlled and accounted for, is an important test before lineshape information can be reliably obtained for other more exotic systems.

In this paper we examine in detail whether ARPES can extract quantitative lineshape information for a model Fermi liquid (FL) system to unambiguously verify the key predictions of the FL description of a correlated material.  The most direct means of testing whether a material is a FL is to determine whether there is a discontinuity in the momentum distribution function $\langle n({\bf k}) \rangle$.  The experimental determination of a discontinuity is fundamentally limited, and may be further complicated in this case by matrix element effects\cite{bansil}.  However, there are a number of directly related signatures that can be tested.  For a FL, Luttinger's counting theorem states that the volume inside the Fermi surface (FS) should be maintained when the interactions are turned on.\cite{luttinger}  Furthermore, if the interactions are not momentum dependent, then the shape of the FS of the correlated material should still be well described by non-interacting calculations.\cite{luttinger-momentum}  Both of these signatures can be directly tested for by ARPES.  

Beyond the volume and shape of the FS, FL theory requires that elementary excitations -- the so called quasiparticles -- close to the Fermi energy have a width that is narrower than their binding energy.  This requirement comes directly from the restricted phase space for low energy scattering processes, and without which the discontinuity in $\langle n({\bf k}) \rangle$ will not occur.  For a 3 dimensional (3D) spherical FS the linewidth is expected to decrease to zero like $\omega^{2}$ as the Fermi energy is approached.\cite{FL}  For a 2 dimensional (2D) cylindrical FS, the dominant term becomes $\omega^{2}|\ln(\omega)|$.\cite{Hodges-2dfl}  The linewidth, and the $\omega$ dependence of the linewidth can, in principle, be measured by ARPES.

TiTe$_{2}$, a correlated material, has been presented in the literature as providing the main example to date of ARPES spectra showing lineshapes well matched by FL theory (see Refs. \onlinecite{claessen}, \onlinecite{perfetti-tite2}, and references within).  However, linewidths in this material have been found to be substantially wider than their binding energy.  This discrepancy has been explained by the 3D nature of TiTe$_{2}$.   The anisotropy in the resistivity between the $ab$ plane and the $c$ direction is only about 40, a small enough difference for electronic dispersion in the $c$ direction to be non-negligible.  With a finite dispersion in the $c$ direction, broadening of the ARPES peaks occurs, including significant final state lifetime effects.\cite{smith-linewidths}  As there is no independent measurement of the $c$ direction dispersion or the photoelectron lifetime in TiTe$_{2}$, the size of these contributions to the overall peak width is not well understood.  

ARPES data from a model FL system would also allow the connection between the transport scattering times and the ARPES derived self-energy to be studied.  In general, this connection is not straightforward as transport measurements and ARPES measure fundamentally different quantities ({\em i.e.} transport does not probe the single particle spectral function).  Furthermore, ARPES measurements are strongly affected by forward scattering events, while transport measurements are dominated by backscattering events.   Other issues include the complication that electron-electron scattering is momentum conserving, and so would not contribute to transport properties unless other scattering processes are available, such as impurity, electron-phonon, and umklapp scattering. 

 TiTe$_{2}$ has also proven to be an interesting case to compare transport results with ARPES spectra.  The temperature dependence of the transport properties implies that the electron-phonon interaction is dominant over a wide temperature range,\cite{allen} which is also inferred by the quasiparticle scattering seen in ARPES.\cite{perfetti-tite2}  Furthermore, the ARPES inferred quasiparticle scattering scales with increasing residual resistivity.\cite{perfetti-tite2}  However, connections beyond these scalings are not possible because the ARPES lineshapes are thought to be strongly effected by the photoelecton lifetime, which is not probed at all by transport measurements.  
  
To date, the best studied ARPES data comes from the surface states found on Cu, Ag, Mo, {\em etc.} where ARPES data interpretation accounts for some of the more subtle effects such as scattering from step edges \cite{felix-1,felix-2}.  However,  surface states are irrelevant for transport properties, and the magnitude of the {\em el-el} coupling is small enough in these materials that it is a challenge to separate it from the more dominant effects close to $E_{f}$.

The difficulties with final state effects in TiTe$_{2}$ and the non-ideal nature of the metal surface states make it necessary to find a strongly 2D system with the  electron-electron interaction as the dominant scattering mechanism in order to investigate FL behavior by ARPES.   \SRO\ has been shown via bulk transport measurements to exhibit good FL behavior below 30 K.\cite{mackenzie-rmp}  It is attractive as a model FL system for ARPES because it can be grown very cleanly and it has a strongly 2D electronic structure\cite{bergemann} with a large electrical anisotropy of about 4000.\cite{mackenzie-rmp}  Previous ARPES measurements on this material have shown a FS\cite{damascelli-sro-bulk} that matches de Haas--van Alphen (dHvA) measurements,\cite{mackenzie-dhva} obeys Luttinger's counting theorem, and is described well by band structure calculations.  

For these reasons \SRO\ is potentially an ideal material to test whether the ARPES linewidths conform to the expectations of FL theory.  Indeed, we find that the intrinsic linewidths of the spectral peaks are narrower than their binding energy and decrease with a functional dependence compatible with FL theory as the Fermi energy is approached.  Furthermore, we find the electron-electron scattering is dominant over the extracted electron-phonon scattering which is consistent with the extended temperature regime (up to 30K) where T$^{2}$ dependent transport behavior is seen.   Equating the self-energy empirically determined in this paper with the transport scattering time of the simple Drude model, we are able to obtain values for  the residual resistivity and the coefficient for the T$^{2}$ temperature dependent part of the resistivity that are in remarkable agreement with the measured transport properties.  These results clearly show that ARPES is a powerful tool to quantitatively study  self-energy effects in highly correlated materials, and in appropriate cases may be able to quantify  momentum dependent self-energy effects.

\section{Experiment}

The surface of \SRO\ has been found to reconstruct when cleaved at low temperatures\cite{matzdorf-sro-surface}.  This altered crystal structure corresponds to an enlarged unit cell which gives rise to an altered electronic structure.  A simplified description of these effects is that ARPES detects three different sets of bands crossing the Fermi energy: the bulk bands, the surface layer bands, and a folded copy of the surface layer bands\cite{damascelli-sro-bulk, shen-sro-surface}(see Fig. 1). Although it has been shown that cleaving at high temperatures suppresses the intensity of the surface related features,\cite{damascelli-sro-bulk} it is far from clear exactly what is happening to the surface for this to occur.  Since we want to gain knowledge about the bulk electronic structure by pursuing a detailed analysis of the ARPES lineshapes -- which are strongly affected by surface degradation -- we have chosen to cleave the sample at low temperature in order to obtain the best quality data.    When this is done, there is one particular location in the Brillioun zone (BZ) that provides enough separation between the surface layer bands and the bulk bands to allow complete analysis of the lineshape of the bulk band.  This location, which gives access to the dispersing  $\alpha$ bands from the surface layer and bulk bands, is in the 2nd BZ along the $(\pi,\pi)$--$(2\pi,\pi)$ cut around ($\frac{4\pi}{3},\pi$), and is indicated as cut ``I'' in Figure  1.  

\begin{figure}
\includegraphics[width=3.4in]{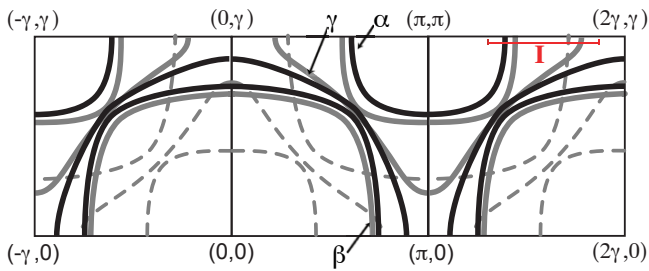}
\caption{Fermi surface schematic of \SRO,\cite{shen-sro-surface} showing the bulk bands (solid black lines) and the surface layer and folded surface layer bands (solid grey lines and dashed grey lines, respectively).  ``I'' indicates the cut of interest for this work.}
\end{figure}

ARPES data were collected on a Scienta-SES200 analyzer at the Stanford Synchrotron Radiation Laboratory Beamline 5-4 with a photon energy of 24 eV, a temperature of 17 K, and a base pressure of $5\times 10^{-11}$ torr.  Additional data were collected on a Scienta-SES2002 analyzer with He I light from a monochromated and modified Gammadata He lamp; the temperature and pressure were 8K and $8\times 10^{-11}$ torr, respectively.  Samples were cleaved {\em in situ} at the measurement temperature.  The energy resolution, assumed to be Gaussian, and determined by fitting Fermi edge spectra from polycrystalline Au was set to either 13 meV.  The angular resolution was measured directly from a point source of electrons traversing a square toothed metallic comb before entering the angular resolved electron detector.  The resulting pattern was best fit with a Gaussian convolution of the original square toothed profile.  The average angular resolution is $0.35^{\circ}$, which corresponds to a momentum resolution of $\approx 0.0126$ \AA$^{-1}$ at these photon energies.

\section{Results}

Working within the Green's function formalism and invoking the sudden approximation, the intensity measured in an ARPES experiment on a 2D single-band material is described by 

\begin{equation}
 I({\bf k},\omega) =[I_0({\bf k},\nu,{\bf A}) f(\omega){\mathcal A}({\bf k},\omega) + B]\otimes R(\Delta {\bf k},\Delta\omega)
\end{equation}

where $I_0({\bf k},\nu,{\bf A})$ is proportional to the one-electron matrix element and dependent on the polarization and energy of the incoming photon, $f(\omega)$ is the Fermi function, and  $B$ is a background. $R(\Delta{\bf k},\Delta\omega)$ is the experimental momentum and energy resolution -- the response function of the instrument. ${\mathcal A}({\bf k},\omega)$ is the single-particle spectral function that contains all the corrections from the many-body  interactions in the form of the self-energy, $\Sigma({\bf k}, \omega)$,

\begin{equation}
  {\mathcal A}({\bf k},\omega)  = \frac{(-1/\pi)\ \ \textrm{Im} \Sigma({\bf k},\omega)}{[\omega - \epsilon^{0}_{k}-\textrm{Re}\Sigma({\bf k},\omega)]^{2}+[\textrm{Im}\Sigma({\bf k},\omega)]^{2}}
 \end{equation}

where $\epsilon^{0}_{k}$ is the bare band dispersion.

Within the Born approximation, the self-energy is written as the sum of individual contributions from all the possible interactions that cause a finite lifetime and energy renormalization of the single particle excitation.  In this work we will be concerned with impurity scattering ({\em imp}), electron-phonon ({\em el-ph}), and electron-electron ({\em el-el}) scattering only. 

The impurity scattering is assumed to be in the strong scattering limit of isotropic point scatters.  The {\em el-ph} interaction will be described by assuming Migdal's theorem is applicable, {\em i.e.}  an isotropic system, a large Fermi surface, and weak to intermediate {\em el-ph} coupling, $\lambda<0.5$.\cite{grimvall}  We will assume the {\em el-el} interaction can be described by FL theory.  Furthermore, we will assume it has negligible ${\bf k}$ dependence, which is consistent with the FS shape being well matched by the non-interacting band structure calculations,\cite{luttinger-momentum} and we neglect any temperature dependence of the {\em el-el} interaction as it is several orders of magnitude smaller than all other terms.   Therefore, the total self energy, $\Sigma_{total}(\omega)$, will be independent of ${\bf k}$ and correspond to  $\Sigma_{imp}+\Sigma_{el\textrm{-}ph}(\omega)+ \Sigma_{el\textrm{-}el}(\omega)$.
 
Since FL theory is normally discussed with relation to transport measurements that only probe quasiparicles within $\pm 2k_{B}T$ of $E_{f}$, it is necessary to justify the much larger energy range (0-60meV) over which we will apply FL theory in this paper.  The initial assumption of FL theory is that only the first order terms in the expansion of the  {\em el-el} self-energy are significant.\cite{FL}  If it is assumed that the coefficients of the higher order terms in the expansion of the {\em el-el} self-energy around $E_{f}$ are of order unity, then they will not become comparable ({\em e.g.} larger than $\approx 1$\% of the leading order terms) until approximately 40--60meV.  

Within the FL framework, the spectral function for a quasi-particle can be rewritten in two slightly different forms which ease analysis.  For the case of $\omega$ very close to $E_f$ , and therefore much smaller than typical phonon energies (although we will not be using an explicit model for the {\em el-ph} interaction, we use the Debye frequency, $\omega_{D} \approx 40meV$ in \SRO\cite{paglione-debye}, as a limiting energy for this case)

\begin{equation}
  {\mathcal A}({\bf k},\omega)  = \frac{(-Z^{total}_k/\pi)\ \ Z^{total}_k\textrm{Im}\Sigma_{total} (\omega)}{[\omega - \epsilon^*_{k} ]^{2}+[Z^{total}_k\textrm{Im} \Sigma_{total} (\omega)]^{2}}
\end{equation}

where we have assumed a linear renormalization for both the {\em el-el} and {\em el-ph} terms so that $Z^{total}_k = (1-\partial \textrm{Re}\Sigma_{el\textrm{-}el} /\partial \omega -\partial \textrm{Re}\Sigma_{el\textrm{-}ph} /\partial \omega ) ^{-1}$ is the total renormalization factor, and $\epsilon^*_{k}= Z^{total}_k\epsilon^{o}_{k}$ is the fully renormalized band energy.
 
If the {\em el-ph} renormalization is not assumed to be linear, as may be the case close to the typical phonon energies, it must be left out of the renormalization factor.  The spectral function can then be written in the following form:
 
 \begin{widetext}
 \begin{equation}
  {\mathcal A}({\bf k},\omega)  =\frac{( -Z^{el\textrm{-}el}_k/\pi)\ \ Z^{el\textrm{-}el}_k\textrm{Im}\Sigma_{total} (\omega)}{[\omega - \epsilon'_{k} -Z^{el\textrm{-}el}_k\textrm{Re}\Sigma_{el\textrm{-}ph}(\omega)]^{2}+[Z^{el\textrm{-}el}_k\textrm{Im} \Sigma_{total} (\omega)]^{2}}
\end{equation}
\end{widetext}
 
 where $Z^{el\textrm{-}el}_k =\left( 1- \partial \textrm{Re}\Sigma_{el\textrm{-}el} /\partial \omega \right) ^{-1}$ is the {\em el-el} renormalization factor, and $\epsilon'_{k} = Z^{el\textrm{-}el}_k\epsilon^{o}_{k}$ is the {\em el-el} renormalized band energy.  
 
Assuming a linear ${\bf k}$ dependence for the renormalized band, either $\epsilon^*_{k} = v^*_{f}{\bf k}$, or  $\epsilon'_{k} = v'_{f}{\bf k}$, it is straightforward within this notation to see what can be determined by the standard energy distribution curve (EDC) and momentum distribution curve (MDC) analysis.   In MDC analysis, a specific $\omega$ value is picked [${\mathcal A}({\bf k},\omega=\textrm{const.})$] and the ${\bf k}$ dependence is that of a Lorentzian.  The peak positions, determined at each value of $\omega$, will generate $ \epsilon^*_{k}/v^*_f $ very close to $E_f$, or more generally $(\epsilon'_{k} -Z^{el\textrm{-}el}_k\textrm{Re}\Sigma_{el\textrm{-}ph}(\omega))/v'_f$.  In both cases the peak width will be  $(2 Z^{el\textrm{-}el}_k\textrm{Im} \Sigma _{total}(\omega))/v'_f = (2 Z^{total}_k\textrm{Im} \Sigma _{total}(\omega))/v^*_f$.  EDC analysis [${\mathcal A}({\bf k}=\textrm{constant},\omega)$], on the other hand, does not allow such immediate identification of the peak position or the peak width.

Note however, that in order to obtain an intrinsic MDC peak position and width from ARPES data, the convolution with the resolution function in Eq. 1,  $R(\Delta {\bf k},\Delta\omega)$, needs to be handled correctly.  The convolution leads to a mixing of ${\bf k}$ and $\omega$, meaning that the shape of the EDCs and MDCs will be affected by both energy and momentum resolution.  In practice, removing the effects of the convolution on a 2D data set [$I(k,\omega)$] is a very challenging problem for which there are two general approaches, as discussed below.   

The first one is to choose an analytic expression, with the smallest number of parameters possible, which is then convolved and fit to the measured $I(k,\omega)$.  Due to the lack of knowledge about the the matrix element, $I_0({\bf k},\nu,\vec{A})$, and the background, B, it is a challenge to guess an appropriate functional form for these terms.  Therefore, applying a full 2D analytical expression, even to a model material, can require a large number of fitting parameters which significantly reduces the overall confidence in the parameters of interest.

The second is to try a direct deconvolution of the numerical data using image analysis procedures.  The deconvolution of ARPES data is unique within the standard set of debluring image analysis problems because we have exact knowledge of our instrument resolution characteristics, or response function, and have a very high signal-to-noise ratio.  Therefore, a simple Wiener filter (see, for example, Ref \onlinecite{numerical_recipes}) can be very effective.  However, if there is a feature in the data that is sharper than the response function, any numerical deconvolution method will be unable to extract it.   In the case of \SRO, the bands are so sharp, as will be shown below, that even 40 meV away from the Fermi function, numerical deconvolution is not an effective tool as it manifests artifacts in the resulting spectra.

Since the two general techniques for deconvolution have serious complications, we propose a third alternative, a combination of a 1D fitting procedure with a 2D convolution of an analytic expression.  By studying the ARPES data as a series of independent EDCs or MDCs (i.e 1D analysis), the direct influences of $I_0({\bf k},\nu,\vec{A})$ on each individual EDC or MDC can, in practice, be neglected.   However, 1D analysis of the 2D data does not allow full energy and momentum broadening to be applied.   We will show that by comparing the 1D analysis of the data with the 1D analysis of a simulation based on a simplified analytical expression that is correctly convolved with $R(\Delta {\bf k},\Delta\omega)$ it is possible to account for the equipment resolution effects while not being hindered by the unknown $I_0({\bf k},\nu,\vec{A})$ and B.

\subsubsection{Bulk $\alpha$ band analysis}

In Fig. 2a, we show the raw data, from the cut marked as ``I'' in Fig. 1.  The surface and bulk $\alpha$ bands on the left hand side of the figure.  In Fig 2d we show just the $\alpha$ bands.  Even though the surface band is relatively weak, but can be clearly distinguished from the bulk band.   The peak position of the MDCs and EDCs have also been added to Fig. 2d for the bulk $\alpha$ band.  One notices immediately that they do not coincide.  The EDC peak position never reaches $E_{f}$, as expected, due to the Fermi function cut off.  It also departs from the MDC peak position at higher binding energies due to effects of the $\textrm{Im}\Sigma(\omega)$.  The larger $\textrm{Im}\Sigma(\omega)$ is, the larger the separation between the MDC and EDC derived peak positions at higher binding energies.  

\begin{figure}
\includegraphics[width=3.4in]{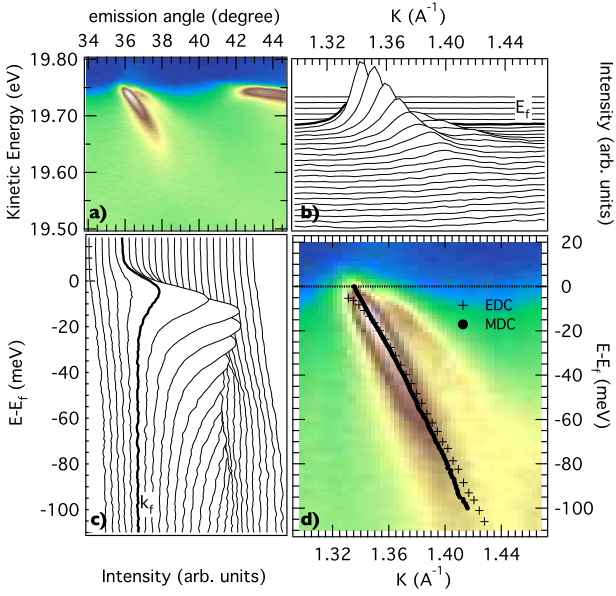}
\caption{(color) The raw ARPES data from cut ``I'' of Figure 1 (a).  The surface and bulk $\alpha$ bands, with the MDC and EDC peak positions of the bulk $\alpha$ band indicated by dots and crosses, respectively (d).  The set of EDCs (b) and MDC's (c) corresponding to the intensity image in d).}
\end{figure}

Figure 3a shows just the MDC peak positions (fitting only the top 30\% of the peak  to minimize location errors from the tails of the MDC peaks) which indicate a subtle kink at $\approx 40$ meV; this matches the Debye temperature of 450 K for \SRO.\cite{paglione-debye}    Also shown in the figure are three straight lines: the unrenormalized band dispersion calculated by LDA, with $v^{0}_{f}= 2.5$ eV\AA\  (grey line);\cite{liebsch} the linearization of the fully renormalized band close to $E_{f}$, $v^*_{f} = 1.02$ eV\AA\ (dotted line); and a fit to the data at $E_{f}$ and beyond 90 meV, which we will take as the {\em el-el} renormalized band, with $v_{f}' = 1.17$ eV\AA\ (solid black line).

\begin{figure}
\includegraphics[width=3.1in]{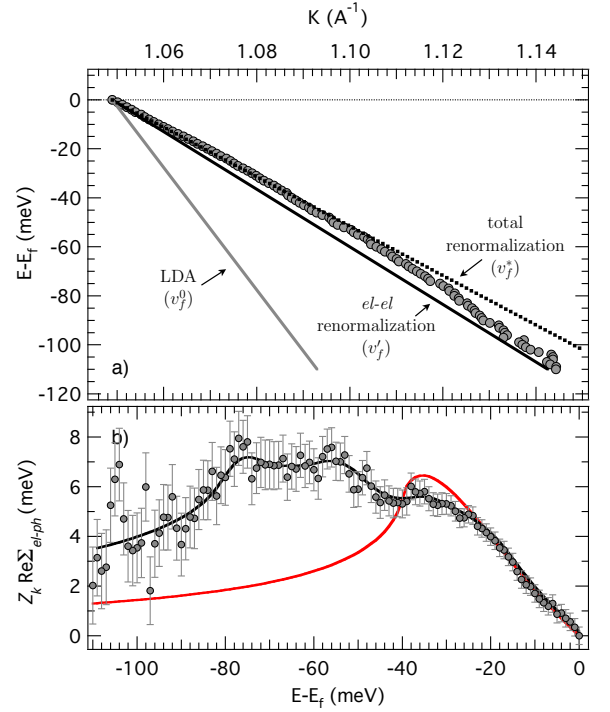}
\caption{a) The MDC derived peak positions of the bulk $\alpha$  band.  The solid grey line is LDA calculation for the bare band, the solid black line is the assumed {\em el-el} renomalized band dispersion (with $v_{f}'= 1.17$ eV\AA), the dot-dash line is fitted to the data from $E_{f}$ to a binding energy of -40 meV, and is used as the fully renormalized band dispersion with $v^*_{f}$ = 1.02 eV\AA.\cite{liebsch}  b) The {\em el-el} renormalized $\Sigma_{el\textrm{-}ph}$ along with a fit to the data, and a fit of the Debye model.}
\end{figure}

$Z^{el\textrm{-}el}_k=0.47$ is calculated from the ratio  $v_{f}'/v^{0}_{f}$; it is assumed to be $\bf k$ independent over the small $\bf k$ range of this particular band.  The ratio  $v^*_{f}/v^{0}_{f}$ gives $Z^{total}_k = 0.41$, which is the total renormalization of the band, and should be compared to the thermodynamic cyclotron mass, $m/m^{*}_{\alpha}= 0.303$, determined by de Hass van Alphen measurements.\cite{mackenzie-dhva, bergemann-2} 

$\lambda$, from the {\em el-ph} mass enhancement factor ($1+\lambda$), is defined as   $- \partial\textrm{Re}\Sigma_{el\textrm{-}ph}(\omega)/\partial\omega|_{\omega = 0}$ and can be determined using $Z^{el\textrm{-}el}_k$ and $Z^{total}_k$ via $\lambda = 1/Z^{total}_k - 1/Z^{el\textrm{-}el}_k$.  It is important to note that $m/m^{*}_{\alpha} \ne (1+\lambda)^{-1}$, and that the size of the kink seen in the data, often given by the ratio of $v^*_f/v'_f$, cannot be used directly to determine $\lambda$, as that will not account for the effect of the  {\em el-el} renormalization.  The value extracted for the {\em el-ph} coupling from this work is $\lambda \approx 0.31$, which is small enough to justify the use of our definition of $\lambda$.\cite{Eiguren}  Data in a recent paper by Y. Aiura {\em et al.}\cite{aiura}  suggests a much larger {\em el-ph} coupling for the $\alpha$ and $\gamma$ bands for data taken at a different point in the BZ.  Beyond the possibility of a strongly $\bf k$ dependent {\em el-ph} coupling, their chosen location in the BZ includes contributions from all three bulk bulk bands, and probably three of the surface layer bands, making it difficult to determine what is contributing to the large kink in their data.

Of primary interest to this work is the width of the peak in MDC analysis, which according to Equation 3 or 4 gives access to  $\textrm{Im}\Sigma_{total}(\omega)$.  However, due to the energy and momentum resolution effects within Equation 1, we do not have direct access to $\textrm{Im}\Sigma_{total}(\omega)$ via the data.  Therefore, we set up a simulation which appropriately accounts for these effects and then directly compare the 1D MDC analysis of the data to that of an identical analysis of the simulated spectra.  If these match we assume we have determined the correct self-energy to describe the data.  The simulation is based on Equations 1 and 4 (setting $I_0=1$ and $B=1$), using the $Z^{el\textrm{-}el}_k$, and $\epsilon'_k = v'_f{\bf k}$ from above, $\textrm{Re}\Sigma_{el\textrm{-}ph}(\omega)$  taken directly from data (and therefore model independent), and $\textrm{Im}\Sigma_{total}(\omega)$ consisting of  $\textrm{Im}\Sigma_{imp}+\textrm{Im}\Sigma_{el\textrm{-}ph}(\omega)+ \textrm{Im}\Sigma_{el\textrm{-}el}(\omega)$.

The functional form of $\textrm{Re}\Sigma_{el\textrm{-}ph}(\omega)$ is determined by the difference between $v'_f$ and the MDC peak position of the data, as shown in Fig. 3b.  The magnitude of $\textrm{Re}\Sigma_{el\textrm{-}ph}(\omega)$ is left as a parameter within the simulation, and is chosen such that the MDC peak positions of the simulated spectra and the data match.    Included in Fig. 3b, for reference, is a fit to a simple Debye model with $\omega_D=40meV$.  There is significant weight above $\omega_D$, suggesting that there may be coupling to multiple phonon modes.  The magnitude of $\textrm{Re}\Sigma_{el\textrm{-}ph}(\omega)$ is chosen so that the MDC peak positions of the simulated spectra and the data match, and so is a free parameter in the simulation.   It should be noted that when the $\textrm{Re}\Sigma_{el\textrm{-}ph}$  starts to decrease (at about 80meV in the data), we can no longer be tracking the quasiparticle, and therefore Equation 4 is no longer strictly valid.\cite{phonon-issues}  

The three terms of $\textrm{Im}\Sigma_{total}(\omega)$ within the simulation are each handled differently.  In the limit of strong scattering from isotropic point scatterers $\textrm{Im}\Sigma_{imp}$ is a constant, which is a free parameter in the simulation.  The functional form of $\textrm{Im}\Sigma_{el\textrm{-}ph}(\omega)$ is determined via the Kramer Kronig transformation of the empirical $\textrm{Re}\Sigma_{el\textrm{-}ph}(\omega)$.\cite{KK-info}   The magnitude of $\textrm{Im}\Sigma_{el\textrm{-}ph}(\omega)$ is fixed by the magnitude chosen for $\textrm{Re}\Sigma_{el\textrm{-}ph}(\omega)$, as discussed above.  The functional form and the magnitude of $\textrm{Im}\Sigma_{el\textrm{-}el}(\omega)$ are left as free parameters in the simulation.

Figure 4 shows the simulation, prior to the convolution with the energy and momentum resolution (a), and after the convolution (b).   The MDC and EDC peak positions of the simulation are also indicated in Figure 4b, and match those of the data when the magnitude of $\textrm{Re}\Sigma_{el\textrm{-}ph}(\omega)$ within the simulation is increased by a factor of 1.3 compared to that found from to data in Figure 3b.  This immediately indicates that even the peak positions in ARPES are influenced by the convolution from the energy and momentum resolution.    Even after this correction, there is one difference between the peak positions in the simulation and the experimental data set,  an up-turn in the MDC peak position as $E_{f}$ is approached.  This is discussed in the appendix.

\begin{figure}
\includegraphics[width=3in]{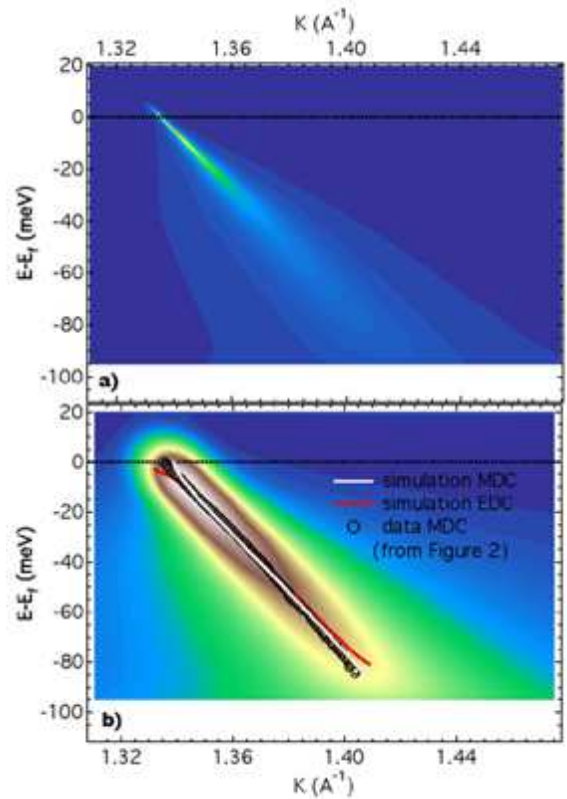}
\caption{(color) Simulated spectra of the bulk $\alpha$ band, with instrument resolution set to zero, $R(\Delta k = 0, \Delta \omega = 0)$ (a), and with $R(\Delta k = 0.0126$\AA$^{-1}, \Delta \omega = 0.013$meV).  The MDC and EDC peak position are indicated by a white and red line, respectively.  Also overlayed on b) is the MDC peak position of the data from Figure 2d.}
\end{figure}

Figure 5 plots the width of the MDC peak as a function of energy of the data from the bulk $\alpha$ band and of several different simulated spectra.  The shape of the MDC peak will not be a Lorentzian, due to the convolution in Equation 1, whenever the energy or momentum resolution is nonzero.  Therefore, to obtain a decent fit to MDC peak shape of the data and the simulated spectra we use a Lorentzian convolved with a Gaussian, whose width is chosen to be the experimental momentum resolution.  The neglected effects of energy resolution within the 1D MDC analysis does not allow the true $\textrm{Im}{\Sigma}(\omega)$ to be extracted, and hence we label the width as $\textrm{Im}\tilde{\Sigma}$.   The substantial MDC width of a simulation where $\textrm{Im}\Sigma_{total}$ only contains $\textrm{Im}\Sigma_{imp}= 1$ meV (the flat black line at  $\textrm{Im}\tilde{\Sigma}\approx 10$ meV in Fig. 5) and $\textrm{Re}\Sigma_{el\textrm{-}ph} = 0$ gives a clear indication of the substantial effects of the neglected energy resolution in this MDC analysis.  Also of significance is the effect of the addition of a $\textrm{Re}\Sigma_{el\textrm{-}ph}$ and $\textrm{Im}\Sigma_{el\textrm{-}ph}$ to the simulation.  By definition  $\textrm{Im}\Sigma_{el\textrm{-}ph}(\omega = 0) =0$, however figure 5 shows an increase in  $\textrm{Im}\tilde{\Sigma}$ at $\omega = 0$.  This is due to the combination of a change in the band position due to  $\textrm{Re}\Sigma_{el\textrm{-}ph}$, over an energy range set by the resolution, and the effects of the full 2D convolution in Equation 1. 

Included in Figure 5 are three simulations where  $\textrm{Im}\Sigma_{imp}+\textrm{Im}\Sigma_{el\textrm{-}ph}(\omega)+ \textrm{Im}\Sigma_{el\textrm{-}el}(\omega)$, but with different functional forms for  $\textrm{Im}\Sigma_{el\textrm{-}el}$.  The blue curve shows  $\textrm{Im}\tilde{\Sigma}$ for a 3D FL form of the {\em el-el} interaction,  $\textrm{Im}\Sigma^{3D}_{el\textrm{-}el} = \beta \omega^2$, with $\beta = 14.5$.  The red curve is for a 2D FL form,\cite{Hodges-2dfl}
 \begin{equation}
\textrm{Im}\Sigma^{2D}_{el\textrm{-}el}(\omega)  =  \beta ' \omega^{2}\left[1+0.53\left|\ln \frac{\omega}{E_{f}} \right|\right],
\end{equation}
where $E_{f}$ is the filled band width, experimentally determined to be approximately 0.5eV for the $\alpha$ band, and $\beta' = 5$.  The best fit to the data over the full 60meV range is found for ${Im}\Sigma^{emp.}_{el\textrm{-}el} = \beta'' \omega^\eta$ with $\beta'' = 2$ and $\eta=1.5$.

\begin{figure}
\includegraphics[width=3.2in]{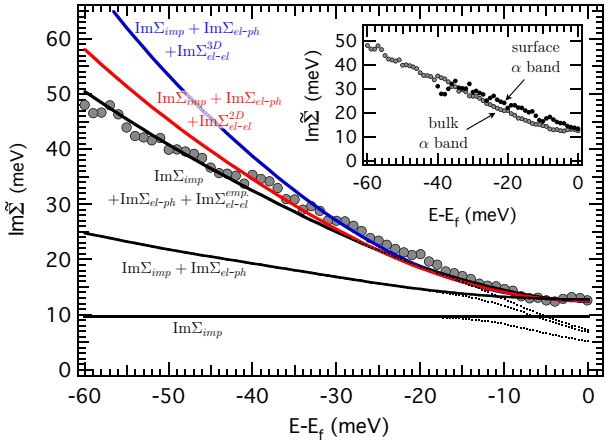}
\caption{(color) $\textrm{Im}\tilde{\Sigma}(\omega)$ as determined by the MDC peak width of the bulk  $\alpha$ band (gray circles). The black lines are $\textrm{Im}\tilde{\Sigma}$ of simulations with only {\em imp} scattering, with {\em imp} and {\em el-ph} scatting, and with three different forms of {\em el-el}.  The blue line is for $\textrm{Im}\Sigma^{3D}_{{el\textrm{-}el}}(\omega)$ ($\beta ' = 14.5$), the red line uses $\textrm{Im}\Sigma^{2D}_{{el\textrm{-}el}}(\omega)$ ($\beta' = 5$), and the black line uses $\textrm{Im}\Sigma^{emp.}_{{el\textrm{-}el}}(\omega)$ ($\beta'' = 2$, $y=1.5$).   The dotted lines that show a decrease in $\textrm{Im}\tilde{\Sigma}$ as $E_f$ is approached are discussed in the appendix.  The inset shows the comparison of $\textrm{Im}\tilde{\Sigma}(\omega)$ for the surface layer $\alpha$ band (black circles) and the bulk $\alpha$ band (grey circles).}

\end{figure}

\subsubsection{Surface layer $\alpha$ band analysis}

Although there is good separation of the bulk and surface layer bands at the BZ location chosen for this work (cut ``I'' in Fig. 1), an attempt at the quantitative analysis on the surface layer $\alpha$ band is problematic.  We find that the noise in the MDC peak positions for the surface layer $\alpha$ band is too large to clearly establish the presence or absence of {\em el-ph} coupling which significantly decreases the confidence level for  determining the functional form of $\textrm{Im}\Sigma_{el\textrm{-}el}(\omega)$.  

However, two qualitative conclusions can be drawn from a comparison  between the extracted $\textrm{Im}\Sigma$ from the MDC width of the surface layer $\alpha$ band compared to that from the bulk $\alpha$ band (shown in the insert of Figure 5).   The first is that the quasiparticles in the surface layer $\alpha$ band have an impurity scattering component to the self-energy that is comparable to that of the bulk states.  Secondly, the comparison between the $\omega$ dependences of the surface layer and bulk $\alpha$ bands suggests that the quasiparticles located in the reconstructed surface layer $\alpha$ band undergo many-body interactions of about the same magnitude as those of the bulk $\alpha$ band. 

\section{Discussion}

With the fitting procedure described above, we are able to retrieve the imaginary part of the self-energy of the bulk $\alpha$ band with minimal {\em a priori} knowledge while accounting for both the instrument energy and momentum resolution.  The extracted self-energy of the bulk band contains three terms:  an isotropic impurity scattering term $\textrm{Im}\Sigma_{imp}  = 1$meV, an {\em el-ph} term defined by $\lambda = 0.31$, and an {\em el-el} term that is best fit by $\textrm{Im}\Sigma= 2\omega^{1.5}$eV.

The self-energy extracted in this work indicates that $\textrm{Im}\Sigma_{total}(\omega)$ approaches zero as we approach the the Fermi energy, implying that at $E_{f}$ the deconvolved MDC and EDC peaks are extremely sharp.  Due to instrument broadening, the raw data will not show this directly. The $\omega$ dependence of $\textrm{Im}\Sigma_{total}$ is best fit over a 60meV window with an {\em el-el} component that suggests scattering with a reduced phase space from the case of a spherical FS of the canonical 3D FL and also from a cylindrical 2D FL ($\approx \omega^{1.67}$).   The $\alpha$ band in \SRO\ is a hybridization of two 1D bands, derived from $d_{xz}$ and $d_{yx}$, which form a very flattened 2D cylindrical FS.  Furthermore there are two other bulk bands (the $\beta$ and $\gamma$ bands) and 6 other surface related bands which cross the Fermi energy.  The phase space present for quasiparticle scattering from the $\alpha$ band, and therefore the functional form of $\textrm{Im}\Sigma$, will be subtly influenced by the coupling between the bands in addition to the shape of the FS.  Therefore, we would not necessarily expect $\textrm{Im}\Sigma_{el\textrm{-}el}(\omega)$ for \SRO\ to have an exact 2D or 3D FL form.  A FL is defined by the presence of quasiparticles, which requires that $\textrm{Im}\Sigma(\omega)$ decreases to zero sufficiently fast with respect to $\textrm{Re}\Sigma(\omega)$ so that there is a discontinuity in $\langle n({\bf k}) \rangle$.  It has recently been carefully shown that the $\omega^2 \ln(\omega)$ dependence of a  2D cylindrical FL does, in fact, show a discontinuity in $\langle n({\bf k}) \rangle$.\cite{halboth}  It is not clear at which point reduced scattering will lead to a breakdown of FL theory, although this work suggests that the 2D cylindrical FS may not be the limiting case.

By assuming that we can make a direct connection between the renormalized imaginary part of the self-energy and the transport scattering time, $1/\tau =  Z_k\textrm{Im}\Sigma/h $,\cite{h_vs_hbar} we can use a simple Drude model ($\rho_{ab}= \Sigma_{i = \alpha, \beta, \gamma} \;m^{*}_{i}/n_{i} e^{2}\tau_{i}$) to draw a comparison with transport properties.  The residual resistivity is therefore determined by the renormalized impurity scattering self-energy ($Z^{total}_k\textrm{Im}\Sigma_{total}(\omega = 0)$).  Furthermore, the $\omega$ dependence of  $\textrm{Im}\Sigma(\omega)_{el\textrm{-}el}$ may be transformed to a temperature dependence -- by replacing $\omega$ with $T$ because the thermal energy has a characteristic energy scale of T from the Fermi energy -- allowing the $T^{2}$ coefficient of the resistivity to be calculated. 

Since we measure $\textrm{Im}\Sigma({\bf k},\omega)$ at only one $\bf k$ point in only one of three bands, we must assume that there is no $\bf k$ dependence, and no band dependence on the scattering rate to calculate $\rho_{ab}$.  Using the mass renormalization from the measured $Z^{total}_k$ of this work for the $\alpha$ band,  the dHvA\cite{mackenzie-dhva} results for the $\beta$ and $\gamma$ bands, and the number of carriers from dHvA\cite{mackenzie-dhva} or the ARPES Fermi surface\cite{damascelli-sro-bulk}, we calculate the residual resistivity to be $\rho_{ab} = 0.2  \mu\Omega$cm.  This value is in very good agreement with the published residual resistivities (0.1 - 0.5 $\mu\Omega$cm) for samples with comparable $T_{c}$.\cite{mackenzie-disorder}   The coefficient of the $T^{2}$ temperature dependent term in the resistivity is calculated\cite{T2_resistivity} to be $5 \times 10^{-3} \mu\Omega$cm K$^{-2}$, very close to the value of $4.5\times 10^{-3} \mu\Omega$cm K$^{-2}$ from Hussey {\em et al.}\cite{hussey}.

Mackenzie et al. \cite{mackenzie-hall} have suggested via the weak field Hall effect that there is no $\bf k $ or band dependence of the impurity scattering.  Therefore the assumptions used to calculate the residual resistivity from the ARPES measured impurity scattering self energy may be appropriate.  This, however, leaves all the possible k-dependence of the self-energy to the electron-electron part.  Ideally photoemission is the perfect tool to study this, and this work shows that with appropriate data handling ARPES can be used to obtain such information. 

Recent work by Kidd {\em et al.}\cite{kidd} attempts to study this issue in \SRO.  They see a lack of significant broadening of the MDC width for the $\beta$ band in \SRO, as a function of temperature or $\omega$, as compared to the $\gamma$ band.  They claim this indicates exotic properties consistent with the quasi-1D nature of the $\beta$ band.  They then use this to argue that this band must play a large role in the high temperature, non-FL, transport properties, while not being significant in the low temperature, FL, transport properties.   In \SRO, the $\alpha$ and $\beta$ bands originate from the $4d_{xz}$ and $4d_{yz}$  Ru orbitals, and are expected to show very similar behavior.  Our work clearly shows that on a very clean surface and with the energy and momentum broadening are handled in detail,  the $\alpha$ band shows very strong $\omega$ dependence in the MDC width.   Also, our calculations of the residual resistivity and the $T^{2}$ temperature dependent coefficient of $\rho_{ab}$ strongly suggest that the quasiparticle scattering from the $\alpha$ band should not be significantly larger than from the $\gamma$ band at the temperature of our experiment.  It is important to note the width of their MDC peaks for both the $\beta$ and $\gamma$ bands at $E_{f}$ are $\approx$ 0.04 \AA$^{-1}$, which is significantly wider that those for the $\alpha$ band from this paper, $\approx$ 0.008 \AA$^{-1}$.   

A note of caution must be mentioned regarding the data analysis of our work.  The initial step in the data analysis, going from Eq. 2 to Eq. 3, specifically makes use of the assumed linear dependence of $\textrm{Re}\Sigma_{el\textrm{-}el}$ on $\omega$.  For a 3D FL, this is appropriate.  However, the logarithmic corrections to $\textrm{Im}\Sigma_{el\textrm{-}el}$ for the 2D case also require corrections to the $\omega$ dependence of $\textrm{Re}\Sigma_{el\textrm{-}el}$.  This is also the case for the {\em emp.} fit.  These nonlinear terms in $\textrm{Re}\Sigma^{2D}_{el\textrm{-}el}$ or  $\textrm{Re}\Sigma^{emp.}_{el\textrm{-}el}$ make the simple interpretation of an MDC as a Lorentzian peak with peak position given by $(\epsilon'_{k} -Z^{el\textrm{-}el}_k\textrm{Re}\Sigma_{el\textrm{-}ph}(\omega))/v'_f$ and a width by $2 Z^{el\textrm{-}el}_k\textrm{Im}\Sigma_{total}/v'_f$ no longer strictly correct. 

\section{Conclusion}

We propose a method to account for the full energy and momentum broadening effects of the  instrument from the ARPES spectra, which then allows quantitative analysis of the quasiparticle many-body interactions.  With this method we find the form of $\textrm{Im}\Sigma(\omega)$ for the  \SRO\ bulk $\alpha$ band to be consistent with a FL: as $E_{f}$ is approached $\textrm{Im}\Sigma \rightarrow \textrm{Im}\Sigma_{imp}$, which we find to be 1meV; and the $\omega$ functional form, best fit by $\textrm{Im}\Sigma^{emp.}_{el\textrm{-}el}(\omega)  =  2 \omega^{1.5}$eV, suggests that the reduced dimensionality is affecting the quasiparticle-quasiparticle scattering.   Beyond the functional form and asymptotic behavior of the extracted $\textrm{Im}\Sigma(\omega)$, the magnitude is also found to be consistent with transport measurements.  Assuming that the quasiparticle lifetime can be directly related to the transport scattering rate, the Drude model can then be used to calculate a residual resistivity in the $ab$ plane, and the $T^{2}$ resistivity coefficient.  Our calculated values ($\rho^{ARPES}_{0} = 0.2  \mu\Omega$cm,  $A^{ARPES} = 5\times 10^{-3} \mu\Omega$cm K$^{-2}$, respectively) are in remarkable agreement with the experimentally determined ones ($\rho^{DC}_{0} \approx 0.1 - 0.5  \mu\Omega$cm, $A^{DC} = 4.5\times 10^{-3} \mu\Omega$cm K$^{-2}$, respectively ).   Furthermore, we find very similar many-body interactions for quasiparticles residing in the surface layer $\alpha$ band.

This work demonstrates that when the full energy and momentum broadening are accounted for, ARPES is capable of showing the expected size of the linewidths for a model Fermi liquid, and gives unique access to the energy dependence of those linewidths.  This analysis, in conjunction with the Fermi surface previously measured by Damascelli {\em et al.}\cite{damascelli-sro-bulk}, give the first complete picture of a Fermi liquid from ARPES.  It also highlights the large effect that energy and momentum broadening can have on measured linewidths, and some of the difficulties in trying to appropriately handle them in the data analysis.  Finally it suggests that ARPES can be a very powerful tool to quantitatively study the $\bf k$ dependence of the self-energy, a concept that is becoming more and more significant.

\begin{acknowledgments}
We thank G.A. Sawatzky for stimulating discussions and acknowledge support from NSF DMR-0304981 and ONR N00014-98-1-0195.  SSRL is operated by the DOE Office of Basic Energy Science under contract DE-AC03-765F00515.
\end{acknowledgments}

\appendix*
\section{Near $E_{f}$ resolution effects}

Within the mathematical model of Eq. 1 the positions of the MDC peaks (see figure 4) of the simulations are strongly altered, turned-up,  at energies close to $E_{f}$ when the energy broadening part of the resolution function, $R(\Delta {\bf k}, \Delta \omega)$,  is larger than the width of the Fermi function ($\Delta \omega > 4k_{B}T$).  This also manifests itself as a strong reduction in the width of the MDC.  In figure 5, for each situation in the graph there are two lines, one of which shows a significant decrease in width as $E_{f}$ is approached starting at a binding energy of $\approx 14$meV.  These dotted lines are calculated from Eq. 1, while the lines that have an asymptotic behavior as $E_{f}$ is approached are calculated with the Fermi function $f(\omega)$ removed from Eq. 1.  

In general, the simplest model of the single particle spectral function, a delta function, when placed in Eq. 1 predicts that both the EDC and MDC peak position are pushed away from the real band position as $E_{f}$ is approached if $\Delta \omega > 4k_{B}T$.  This fact makes it clear that the effect is not related to the presence, or form, of the self-energy, but is rather {\em directly connected to the convolution of the Fermi function with the resolution function}.  

This predicted MDC turn-up has not been seen in the published \SRO\ data \cite{aiura,kidd,shen-sro-surface}, nor is it seen in the published cuprate ARPES data [see Ref. \onlinecite{ad-review}, and references within].  It has also not been seen in the surface state data on Cu, Ag, Au,\cite{reinert-ag-au-cu}, or the quantum well states in Pb on Si,\cite{upton}, however these data are usually taken with resolution and temperature conditions such that $ 4k_{B}T >\Delta \omega $.  It should be noted that there will be a shift in $k_{f}$, as determined by MDC analysis, with respect to the true $k_f$ when a finite resolution is used in Eq. 1.

In an attempt to understand this issue, it is worth looking at both the presence of the Fermi function and the form of the resolution broadening.   The Fermi function is present in Equation 1 in order to introduce a temperature dependence to a zero temperature single particle spectral function, ${\mathcal A}(k,\omega)$.  In a noninteracting system there is no temperature dependence of ${\mathcal A}(k,\omega)$. However, ARPES only measures the $N-1$, or electron removal, part of ${\mathcal A}(k,\omega)$ which does have a temperature dependence given by the Fermi function.  As interactions are adiabatically turned on, this will continue to hold.  Therefore, in the case of a FL, such as \SRO, we expect that ${\mathcal A}(k,\omega) f(\omega)$ is the correct form to include temperature dependence.  For a non-FL spectral function (such as that expected for the cuprates) this argument, in general, will not hold.  

Experimentally, the resolution function may not be appropriately accounted for by the convolution in Equation 1.  We can directly measure the momentum broadening, as mentioned in the Experimental section, and find that it does look like a gaussian convolution.  However, we do not have such direct access to the appropriate functional form of the energy broadening.   At present, an unexplained effect of the resolution function looks like the most likely culprit for the discrepancy close to $E_f$ between the simulation via Equation 1 and the data.


\begin{thebibliography}{36}
\bibitem[{pho()}]{phonon-issues}
\bibinfo{howpublished}{The concept of a well defined excitation spectrum, a
  quasiparticle, fails severely when the electronic excitation is of the order
  of a typical phonon energy -- there is no longer a simple pole structure in
  the spectral function. An indicator that there are problems with the
  interpretation of equation 4 at energies greater than the typical phonon
  energies is that the quasiparticle dispersion, given by the MDC peak
  position, can give an effective mass that is smaller than that of the bare
  quasiparticle. See Ref. 17 for a further discussion of this issue.}
\bibitem[{KK-()}]{KK-info}
\bibinfo{howpublished}{In order to numerically calculate the Kramers Kronig
  transformation the $\textrm{Re}\Sigma_{el\textrm{-}ph}$ was analytically
  continued beyond 110 meV by assuming an asymptotic decrease to zero. The
  functional form of the continuation has only a minor effect on the transform
  in the 0 - 60 meV range}.
\bibitem[{h_v()}]{h_vs_hbar}
\bibinfo{howpublished}{In the Drude model the scattering time is a true time,
  related to a true frequency, $\nu$, not an angular frequency $\omega$.
  Therefore we use $h$ instead of $\hbar$ to convert $\textrm{Im}\Sigma$ into a
  scattering time.}
\bibitem[{T2_()}]{T2_resistivity}
\bibinfo{howpublished}{We have used the $\beta$ coefficient for the 3D FL case
  ($\beta = 14.5$) to make the comparison with the $T^2$ term from transport}.
\end{thebibliography}
\end{document}